\documentclass[12pt]{article}

\usepackage{sbc-template}
\usepackage{graphicx,url}
\usepackage[utf8]{inputenc}
\usepackage[brazil]{babel}
\usepackage{adjustbox}
\usepackage{tikz}
    
\title{The professional's opinion: Suggestions for improving the corporate education training process in Software Engineering.}

\author{Rodrigo Siqueira\inst{1}, Danilo Monteiro Ribeiro\inst{1} }

\address{CESAR School \\  Recife -- PE -- Brazil}

\begin{document}

\maketitle

\begin{abstract}
Technology organizations continuously invest in professional development, but face difficulties in transferring learning to project practice. This exploratory qualitative study investigates which improvements software engineering professionals suggest for organizational learning processes. 174 open-ended responses were analyzed through reflexive thematic analysis. Five themes emerged: practical applicability and alignment with needs; pedagogical quality and organization; time and structural conditions; incentives and institutional recognition; and interaction, mentoring, and social exchange. The results indicate that improving learning requires systemic interventions that integrate practical relevance, structural support, and a favorable institutional culture.
\end{abstract}

\section{Introduction}

The acquisition of knowledge, skills, and attitudes for current tasks is a central element for individuals to contribute to their organizations \cite{fitzgerald1992training}. Research on training consistently highlights two findings: (a) training is effective and (b) the design, delivery, and implementation of programs significantly influence their outcomes \cite{salas2012science}. In contemporary organizations, continuous learning has become imperative, and companies invest in training because they recognize that a skilled and constantly evolving workforce constitutes a strategic advantage \cite{salas2012science}.

This need is particularly pronounced in the information technology sector. IT companies face pressure to build and maintain high-quality human capital, essential for sustaining innovation and competitiveness \cite{diniz2024skill}. The sector is characterized as a training-intensive market, in which training is an integral part of the operational environment \cite{zornada2005learning}. A mismatch between the offerings of educational institutions and the demands of the software industry creates a shortage of adequately prepared professionals \cite{diniz2024skill}, forcing companies to invest in internal programs to bridge this competency gap \cite{diniz2024skill}. Training and development activities are implemented to enhance employees' competencies in alignment with organizational objectives \cite{lee2025overcoming}.

Despite these investments, the effective transfer of learning to the work context remains a challenge. The literature positions training as a relevant component in cultural change processes and as a determinant of organizational performance \cite{santos2003employee}. However, although professionals acknowledge the value of training, this perception weakens when one questions whether high investments actually result in adequate technical competencies \cite{ragnarsdottir2011training}. Many organizations lack mechanisms to evaluate the effectiveness of their programs \cite{santos2003employee}, and determining the return on investment in technology training remains an open problem \cite{coverstone2003training}.

The lack of focus on job- and task-level needs can compromise the relevance of training initiatives, since the identification of specific competencies constitutes a starting point for effective programs \cite{de2025mapping}. It is recommended that organizations regularly monitor employee satisfaction \cite{velada2007training}, considering that satisfied individuals show a greater propensity to transfer learning \cite{velada2007training}. Collecting feedback through surveys and interviews allows for examining perspectives on the value of training and preferred formats \cite{lee2025overcoming}, and such evaluations provide input for future program modifications \cite{salas2012science}. Despite these recommendations, qualitative studies investigating improvement suggestions from the perspective of software engineering professionals themselves remain scarce.

Given this gap, this study seeks to answer the following research question:

\begin{quote}
    \textbf{RQ1 --- What are the main improvement suggestions made by software engineering professionals to enhance the learning process in corporate education?}
\end{quote}

This study contributes by offering a qualitative perspective on the improvements perceived as necessary in organizational learning processes in software engineering, complementing existing findings and providing input for the design of training programs better aligned with professionals' needs.

The remainder of this paper is organized as follows: Section~\ref{sec:background} presents the theoretical background; Section~\ref{sec:metodo} describes the methodology; Section~\ref{sec:resultados} presents the results; Section~\ref{sec:discussao} discusses the findings and their implications; Section~\ref{sec:limitacoes} addresses limitations and threats to validity; and Section~\ref{sec:conclusao} concludes the paper.

\section{Background}
\label{sec:background}

Training success is not treated as an isolated event, but as a process involving antecedent factors, dynamics during instruction, and subsequent conditions \cite{salas2012science}. Enrolling employees in training programs within unfavorable organizational contexts can result in a waste of development resources \cite{santos2003employee}.

The practical application of learning remains limited. It is estimated that approximately 10\% of the content learned in training sessions is effectively applied at work \cite{sofo2007transfer, fitzpatrick2001strange}. Resistance tends to increase when content is perceived as irrelevant \cite{kupritz2002relative}, and experienced professionals may evaluate certain trainings as superfluous \cite{ohlmann2019perception}.

Training Needs Analysis (TNA), through task analysis, aims to specify critical job functions and identify required knowledge, skills, and attitudes \cite{salas2012science}. Cognitive task analysis complements this approach by understanding the requirements and mental processing involved in occupational performance \cite{salas2001science}. Both serve to align instructional content with job demands.

Pedagogical quality is also highlighted. Effective training provides structured opportunities for learning through instruction, demonstration, practice, and timely diagnostic feedback \cite{salas2001science}. The instructor is considered central to program effectiveness \cite{fawad2012integrated}, with a particularly relevant impact in technical courses \cite{devaraj2004measure}.

The literature indicates that learning and its transfer depend on both the individual's capability and motivation \cite{santos2003employee}. The opportunity to perform newly learned behaviors at work is a necessary condition for consolidating transfer \cite{quinones1995effects}. The absence of practical examples and interactivity in the work context is a limiting factor \cite{salas2012science}.

Organizational support is a relevant variable. Professionals who perceive encouragement from their managers are more likely to participate in training (58\%) compared to those with little or no encouragement (24\%) \cite{santos2003employee}. Additionally, 53\% of respondents in one study reported not being held accountable for demonstrating competencies after training \cite{coverstone2003training}. Performance evaluation and personal development planning systems can establish links between development, career progression, and rewards \cite{santos2003employee}.

Regarding the social dimension, variables such as opportunity to use, peer support, and supervisor support influence post-training behavior and transfer motivation \cite{seyler1998factors}. Coworkers constitute a relevant source of knowledge, with a positive perception of skill sharing \cite{ragnarsdottir2011training}. The literature emphasizes the role of leaders as development facilitators, including coaching and debriefings \cite{salas2012science}.

In summary, training effectiveness depends on an articulated set of factors related to instructional design, individual characteristics, and the environmental conditions that support the application of learning.

\section{Method}
\label{sec:metodo}

\subsection{Research Design}

This study is characterized as exploratory qualitative research, grounded in Reflexive Thematic Analysis \cite{terry2017thematic}. The analyzed question was extracted from a broader survey, approved by the Research Ethics Committee (CAEE: 91121125.4.0000.5208; Report No. 7.816.810). This work focuses on the following open-ended question, whose objective was to understand the improvements perceived by professionals in software engineering learning processes:

\begin{quote}
``\textbf{If you could suggest one improvement to the learning processes, what would it be?}''
\end{quote}

\subsection{Participants and Context}

The survey included 283 participants. For this qualitative stage, \textbf{174 valid responses} to the open-ended question were considered. This reduction is due to the non-mandatory nature of the open-ended question and the exclusion of entries without analytical content, such as ``no comments,'' ``I don't know,'' or ``I have no suggestions.''

The sociodemographic profile of respondents indicates a predominance of males (74.7\%), with a mean age of 33.9 years. Geographic concentration was observed in Pernambuco (39.1\%) and São Paulo (20.7\%). Regarding education, 31.0\% hold a completed undergraduate degree and 58.6\% hold a graduate degree (completed or in progress). The majority work in companies with 100 or more employees (78.2\%). The predominant area is Software Development (41.4\%), followed by People/Project Management (16.1\%) and Quality Assurance (15.5\%). The sample exhibits high seniority, with concentration in Senior (31.6\%), Mid-level (20.7\%), Manager (18.4\%), and Specialist (14.9\%) positions, with 44.8\% having more than eight years of experience in Software Engineering.

\subsection{Data Collection}

Data collection was conducted remotely and asynchronously through a questionnaire on the Google Forms platform. Recruitment was carried out through professional social networks, particularly LinkedIn, as well as groups on messaging applications. Participation was voluntary, with consent provided through a digital Informed Consent Form.

\subsection{Analysis Procedure}

Data were analyzed using Reflexive Thematic Analysis \cite{terry2017thematic}, following six phases. The first consisted of \textbf{familiarization} with the data, through a thorough reading of the responses and recording of preliminary notes. In \textbf{initial coding}, relevant excerpts were segmented and associated with descriptive codes. \textbf{Theme construction} involved examining codes for convergence patterns and organizing them into meaning clusters. In the \textbf{review and refinement} phase, the internal coherence of clusters and the distinction between categories were verified. \textbf{Definition and naming} comprised the conceptual delimitation and final naming of themes. Finally, the \textbf{report production} involved the narrative integration of themes into the body of the paper.

Semantic consistency was maintained and overlap between categories was avoided. When a single participant used multiple expressions related to a single pattern, such occurrences were grouped within the overall interpretive context. To illustrate this process, Table~\ref{tab:exemplo_analise} follows a single response across the phases, demonstrating how raw data were progressively segmented into codes, grouped into initial themes, and consolidated into final themes.

\begin{table}[htbp]
    \centering
    \caption{Example of the thematic analysis and co-occurrence analysis process applied to a single response}
    \label{tab:exemplo_analise}
    \begin{tabular}{|l|p{10cm}|}
        \hline
        \textbf{Phase} & \textbf{Result} \\
        \hline
        \textbf{1 -- Familiarization} &
        P173: ``I would suggest integrating learning into the daily routine, with regular exchanges, mentoring, and real case studies. That way, knowledge wouldn't be restricted to one-off trainings but would become part of everyday work.'' \\
        \hline
        \textbf{2 -- Initial coding} &
        \textit{Routine}; \textit{Mentoring}; \textit{Real cases}; \textit{Everyday work} \\
        \hline
        \textbf{3 -- Theme construction} &
        \textit{Practical applicability of content}; \textit{Social interaction and experience sharing}. \\
        \hline
        \textbf{4 -- Review and refinement} &
        Verification of internal coherence: the codes \textit{Routine}, \textit{Real cases}, and \textit{Everyday work} were consolidated under the practical applicability axis; \textit{Mentoring} was associated with the social interaction dimension. \\
        \hline
        \textbf{5 -- Definition and naming} &
        \textbf{T1} -- Practical applicability and alignment with needs; \textbf{T5} -- Interaction, mentoring, and social exchange. \\
        \hline
        \textbf{6 -- Report production} &
        Narrative integration of themes into the paper, with selection of illustrative quotes and articulation with the literature. \\
        \hline
        \textbf{Co-occurrence} &
        This response was assigned two themes (\textbf{T1} and \textbf{T5}), generating a co-occurrence record between them. The sum of these records across the entire corpus comprises the edges and weights of the graph in Figure~\ref{fig:network_temas}. \\
        \hline
    \end{tabular}
\end{table}

\section{Results}
\label{sec:resultados}

This section presents the findings of the thematic analysis, whose objective was to identify opportunities for improving organizational training models from the perspective of software engineering professionals.

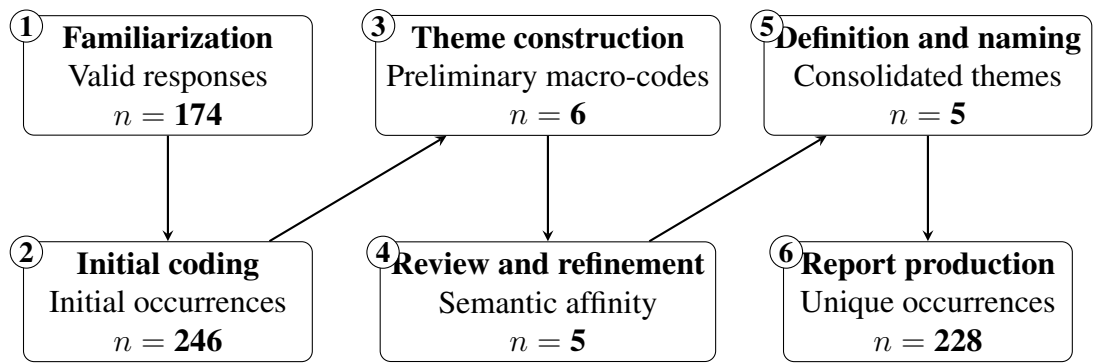
\begin{figure}[htbp]
    \centering
    \tikzset{
        phase/.style={rectangle, draw, align=center, minimum height=1.5cm, minimum width=3.8cm, rounded corners, fill=white},
        number/.style={circle, draw, fill=white, inner sep=1pt, minimum size=0.4cm, font=\small\bfseries},
        arrow/.style={thick, ->, >=stealth}
    }
    \begin{adjustbox}{max width=\textwidth}
        \begin{tikzpicture}
            \node[phase] (p1) at (0, 0) {\textbf{Familiarization} \\ Valid responses \\ $n=\textbf{174}$};
            \node[phase] (p2) at (0, -3) {\textbf{Initial coding} \\ Initial occurrences \\ $n=\textbf{246}$};
            \node[phase] (p3) at (5, 0) {\textbf{Theme construction} \\ Preliminary macro-codes \\ $n=\textbf{6}$};
            \node[phase] (p4) at (5, -3) {\textbf{Review and refinement} \\ Semantic affinity \\ $n=\textbf{5}$};
            \node[phase] (p5) at (10, 0) {\textbf{Definition and naming} \\ Consolidated themes \\ $n=\textbf{5}$};
            \node[phase] (p6) at (10, -3) {\textbf{Report production} \\ Unique occurrences \\ $n=\textbf{228}$};
            \node[number, anchor=south east, xshift=.2cm, yshift=-.3cm] at (p1.north west) {1};
            \node[number, anchor=south east, xshift=.2cm, yshift=-.3cm] at (p2.north west) {2};
            \node[number, anchor=south east, xshift=.2cm, yshift=-.3cm] at (p3.north west) {3};
            \node[number, anchor=south east, xshift=.2cm, yshift=-.3cm] at (p4.north west) {4};
            \node[number, anchor=south east, xshift=.2cm, yshift=-.3cm] at (p5.north west) {5};
            \node[number, anchor=south east, xshift=.2cm, yshift=-.3cm] at (p6.north west) {6};
            \draw[arrow] (p1) -- (p2);
            \draw[arrow] (p2) -- (p3);
            \draw[arrow] (p3) -- (p4);
            \draw[arrow] (p4) -- (p5);
            \draw[arrow] (p5) -- (p6);
        \end{tikzpicture}
    \end{adjustbox}
    \caption{Phases of thematic analysis and refinement quantitative at each stage}
    \label{fig:phases_tematica}
\end{figure}

In \textbf{Phase 1 (familiarization)}, the entire textual corpus was read. In \textbf{Phase 2 (initial coding)}, fragments were segmented into units of meaning, generating 246 coding occurrences. In \textbf{Phase 3 (theme construction)}, these codes were grouped into 6 macro-codes. In \textbf{Phases 4 and 5 (review, refinement, definition, and naming)}, macro-codes with strong semantic affinity were grouped into a single thematic axis. To avoid over-representation, multiple synonymous terms from the same participant belonging to the same theme were grouped into a single instance. \textbf{Five main themes} were consolidated, encompassing \textbf{228 unique thematic occurrences}. The final frequency exceeding the number of participants (174) reflects the multifaceted nature of the suggestions.

\begin{table}[htbp]
    \centering
    \caption{Frequency of identified themes}
    \label{tab:frequencia_temas_qq2}
    \begin{tabular}{|l|p{8cm}|c|}
        \hline
        \textbf{Code} & \textbf{Theme} & \textbf{Frequency} \\
        \hline
        T1 & Practical applicability and alignment with needs & 87 \\
        T2 & Pedagogical quality, didactics, and organization & 70 \\
        T3 & Time and structural conditions for learning & 28 \\
        T4 & Incentives, recognition, and institutional support & 27 \\
        T5 & Interaction, mentoring, and social exchange & 16 \\
        \hline
    \end{tabular}
\end{table}

The predominant demand is concentrated in T1 ($n = 87$), followed by T2 ($n = 70$). Professionals seek training pathways that are meaningful for their daily activities, respecting the context and stage of each career.

\begin{figure}[htbp]
    \centering
    \includegraphics[width=0.9\textwidth]{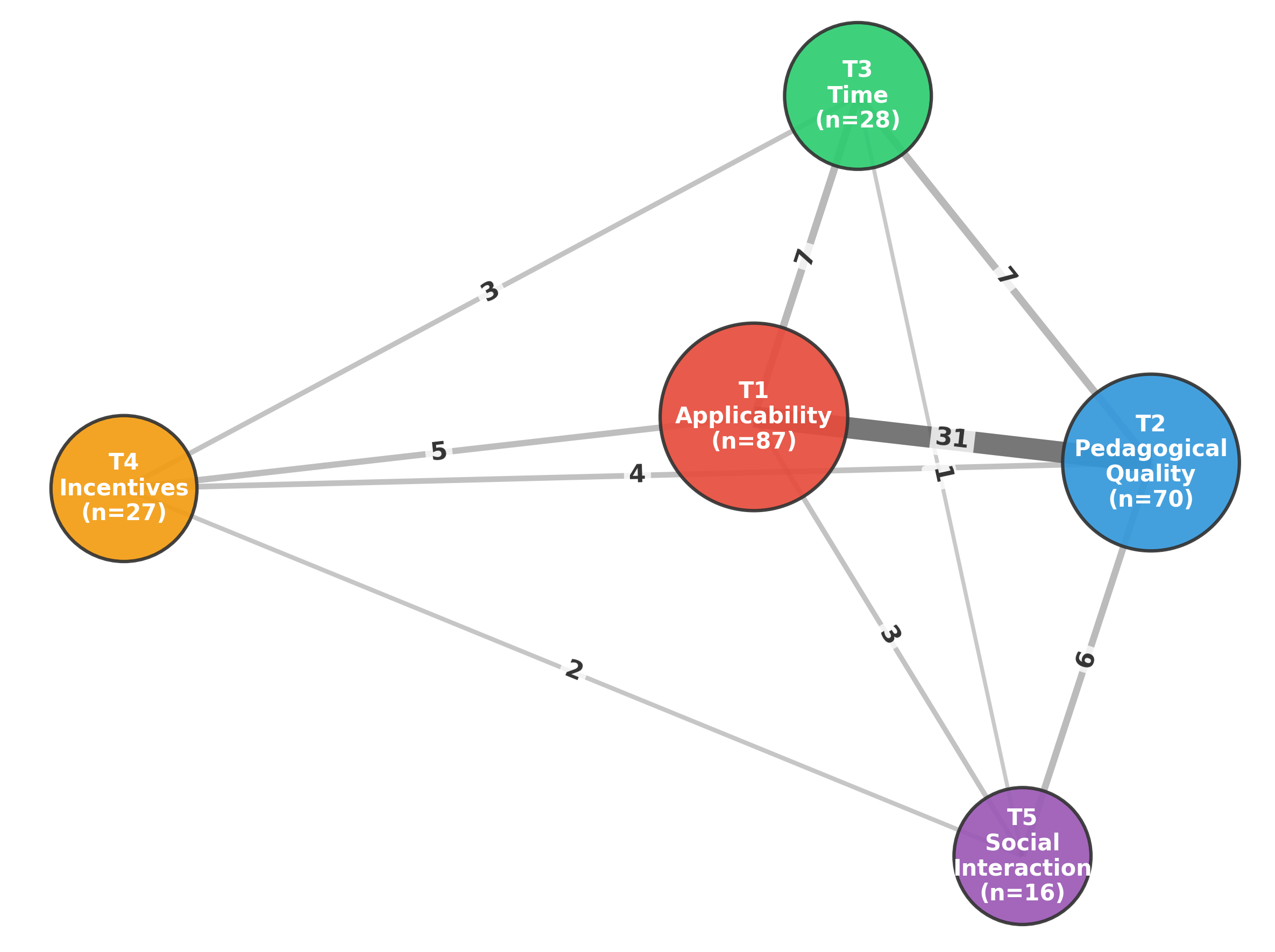}
    \caption{Co-occurrence graph between identified themes.}
    \label{fig:network_temas}
    \par\vspace{4pt}
    {\small
    \begin{tabular}{@{}rl@{}}
        (1) & Each node represents a theme, with its frequency in parentheses. \\
        (2) & Each edge connects themes cited in the same response. \\
        (3) & The number on the edge indicates the number of co-occurrences. \\
        (4) & Edge thickness is proportional to the number of co-occurrences. \\
    \end{tabular}}
\end{figure}

Beyond isolated frequency, co-occurrence between themes was analyzed, that is, how often two themes were cited simultaneously by the same participant. This analysis was possible because each response could contain suggestions related to more than one theme. Figure~\ref{fig:network_temas} represents this relationship. A strong core is observed between \textbf{T1} and \textbf{T2}, with 31 co-occurrences, indicating that the practical utility of content is inseparable from efficient didactic structuring. Infrastructure themes, such as \textbf{T3}, orbit this core, suggesting that temporal viability is a critical enabler for the articulation between theory, practice, and context to occur in the organizational environment.

Below, each theme is detailed based on qualitative evidence from the responses. To preserve anonymity, direct quotes are identified with the prefix ``P'' followed by the participant number.

\subsection{T1 -- Practical applicability and alignment with needs}

This theme encompasses most of the suggestions and reflects what is learned and how it is used. Participants emphasize that training only fulfills its purpose when focused on hands-on activities, real problems, and everyday simulations. It also includes responses that emphasize alignment with career and project objectives.

\begin{quote}
\textit{P5: I would suggest that learning processes be more based on real practice. We often receive theoretical training, but we only retain it when we apply it in our daily work. If training programs included simulations, practical labs, or real project challenges.}
\end{quote}

\begin{quote}
\textit{P9: Be more connected with the employee's daily work, not only with the company's strategy.}
\end{quote}

\subsection{T2 -- Pedagogical quality, didactics, and organization}

This theme concentrates suggestions about content quality, instructor didactics, class dynamism, and support materials. Participants demand less generic training and highlight the importance of the preparation of those who lead the training.

\begin{quote}
\textit{P88: Instructors with better didactics and more up-to-date materials.}
\end{quote}

\begin{quote}
\textit{P18: Most classes are disconnected and superficial on topics that do not raise the level of technical and research knowledge of the students.}
\end{quote}

\begin{quote}
\textit{P177: Creating documentation to guide learning is always useful, as it provides clarity about the path to follow. Learning 'in the dark' has its value, but having a guide makes the process much more efficient.}
\end{quote}

\subsection{T3 -- Time and structural conditions for learning}

This theme indicates the need for protected time during work hours for study and practical application, without competition from operational demands.

\begin{quote}
\textit{P47: Even though the company encourages participation, project deadlines remain the same and there are always important meetings that cannot be rescheduled. Ideally, there would be reserved time for study, where only studying would be the focus during that time.}
\end{quote}

\begin{quote}
\textit{P49: Allow more work hours to be used for dedicating to learning processes.}
\end{quote}

\subsection{T4 -- Incentives, recognition, and institutional support}

This theme covers responses about the organization's role in funding education, providing learning tool support, and recognizing the effort of those who pursue continuous training.

\begin{quote}
\textit{P4: I prefer that the company pay for a school and provide access to employees rather than create something locally on its own.}
\end{quote}

\begin{quote}
\textit{P40: Support for participation in these events.}
\end{quote}

\begin{quote}
\textit{P43: I believe that companies sometimes only provide financial and/or time incentives (by releasing the employee) but do not recognize the effort nor present it to the company as a whole so that it can serve as an incentive for other colleagues.}
\end{quote}

\subsection{T5 -- Interaction, mentoring, and social exchange}

This theme gathers responses that highlight the importance of interactions with leaders, senior mentors, and peers, as well as forums and smaller class sizes. It addresses the relational dimension of learning.

\begin{quote}
\textit{P1: More interaction with instructors and smaller class sizes.}
\end{quote}

\begin{quote}
\textit{P17: Discussion forums with moderators who had a script to guide but that didn't seem too rigid.}
\end{quote}

\begin{quote}
\textit{P61: A less experienced developer could be responsible for investigating and proposing the solution, with the support of a senior professional.}
\end{quote}

Subthemes reveal specific dimensions within each central axis. In T1, hands-on learning and alignment with career objectives stand out. In T2, the focus is on didactic effectiveness and content. T3 is subdivided between protected hours and balance with operational workload. T4 focuses on both financial support and a culture of recognition. T5 branches into technical mentoring and collaborative exchange spaces among peers.

The results indicate that the suggested improvements go beyond the provision of technical content. The demands require systemic interventions: structural feasibility, didactic improvement, and contextualization of learning for project practice.

\section{Discussion}
\label{sec:discussao}

The results reveal that organizational learning in Software Engineering is perceived as a multidimensional phenomenon. The five identified themes converge with the understanding that training success involves antecedent factors, instructional dynamics, and subsequent conditions \cite{salas2012science}. The emergence of 228 thematic occurrences from 174 respondents reinforces this systemic perspective \cite{salas2012science}.

The theme \textit{Practical applicability} ($n = 87$) concentrated most of the suggestions. Participants call for simulations, labs, and challenges based on real everyday problems, signaling that training is only perceived as effective when connected to professional practice. This demand is consistent with evidence that the opportunity to perform newly learned behaviors is a necessary condition for transfer \cite{quinones1995effects}, and that the absence of practical examples limits instructional effectiveness \cite{salas2012science}. The senior profile of the sample may amplify this perception, given that experienced professionals tend to evaluate decontextualized trainings as superfluous \cite{ohlmann2019perception} and that resistance increases in the face of irrelevant content \cite{kupritz2002relative}. These findings reinforce the relevance of Training Needs Analysis \cite{salas2012science, salas2001science}.

\textit{Pedagogical quality} ($n = 70$) shows that the way content is structured directly influences engagement. Participants reported dissatisfaction with generic training and poorly prepared instructors. The literature supports that effective training requires instruction, demonstration, practice, and timely feedback \cite{salas2001science}, and that instructor skill is a critical variable in technical courses \cite{fawad2012integrated, devaraj2004measure}. The proximity between T1 and T2 suggests that applicability and pedagogical quality are complementary faces of the same requirement. The network graph (Figure~\ref{fig:network_temas}) ratifies this ``value core'': the utility of content (the ``what'') is linked to the effectiveness of delivery (the ``how''). Failures in pedagogical delivery can compromise the perception of applicability regardless of technical relevance.

The theme \textit{Time and structural conditions} ($n = 28$) expresses a tension between the discourse of encouraging training and the concrete conditions for its realization. Reports describe scenarios in which deadlines and operational demands compete with study time, showing that providing training resources is insufficient without structural conditions for their utilization. The literature corroborates this by indicating that training employees in unfavorable contexts can result in a waste of resources \cite{santos2003employee}.

\textit{Incentives and institutional support} ($n = 27$) situate learning within organizational culture. Financial investment is not sufficient without recognition of the training effort. This perception is consistent with evidence that managerial encouragement increases training participation \cite{santos2003employee} and that the absence of accountability weakens engagement \cite{coverstone2003training}. Systems that link development, career, and rewards are recommended \cite{santos2003employee, velada2007training}.

Although \textit{Interaction, mentoring, and social exchange} showed the lowest frequency ($n = 16$), its presence illuminates the relational dimension of learning. Variables such as peer support and supervisor support influence post-training behavior \cite{seyler1998factors}, and coworkers are recognized as a relevant source of knowledge \cite{ragnarsdottir2011training}. The lower frequency may reflect professionals' tendency to prioritize more immediately perceived gaps, while the social dimension acts as a substrate that enhances the others.

The five themes indicate that learning effectiveness depends on systemic integration among its dimensions. Transfer failures may stem less from content quality and more from systemic misalignments. Based on the identified patterns, the following analytical propositions are derived for future investigations:

\begin{itemize}
    \item \textbf{PR1:} Practical applicability constitutes the central dimension in the perception of learning effectiveness.
    \item \textbf{PR2:} Instructional quality influences motivation for participation in training programs.
    \item \textbf{PR3:} Protected time acts as a moderator of learning transfer.
    \item \textbf{PR4:} Institutional recognition impacts continuous engagement in training processes.
    \item \textbf{PR5:} Social interaction enhances the retention and application of acquired knowledge.
\end{itemize}

These propositions do not claim confirmatory status, but offer directions for quantitative studies seeking to validate the relationships suggested here.

\section{Limitations and Threats to Validity}
\label{sec:limitacoes}

This study has limitations that should be considered. The potential influence of researchers on the qualitative data analysis is acknowledged, inherent to the reflexive thematic approach. The use of a single open-ended question, although allowing free expression by participants, limits the depth and breadth of responses when compared to techniques such as semi-structured interviews or focus groups. The sample is restricted to professionals recruited in Brazil, limiting generalization. The high educational profile (88.3\% with completed higher education or graduate degrees) may not reflect the entirety of the sector's workforce. Professional level categories may have been interpreted heterogeneously. The findings are circumscribed to software engineering. Future investigations should prioritize longitudinal studies, cross-cultural comparisons, objective effectiveness metrics, and the exploration of sociodemographic variables.

\section{Conclusion}
\label{sec:conclusao}

Organizational learning is a strategic factor in software engineering. Investments in training aim to enhance human capital, contributing to productivity, process optimization, and talent retention \cite{lee2025overcoming}. However, a gap persists between what is taught and what is applied in daily practice, and qualitative studies from the professionals' perspective remain scarce \cite{de2025mapping}.

This study identified five central themes from 174 responses analyzed via Reflexive Thematic Analysis \cite{terry2017thematic}: (T1) Practical applicability ($n = 87$), (T2) Pedagogical quality ($n = 70$), (T3) Time and structural conditions ($n = 28$), (T4) Incentives and institutional support ($n = 27$), and (T5) Interaction, mentoring, and social exchange ($n = 16$). These findings reinforce that training should be understood as a system, in which what occurs before, during, and after training is determinant for its effectiveness \cite{salas2012science}.

In response to \textbf{RQ1, what are the main improvement suggestions made by software engineering professionals to enhance the learning process in corporate education}, the results indicate five priority dimensions. Practical applicability emerged as the central dimension, signaling that decontextualized programs tend to generate resistance and low transfer. Pedagogical quality and instructor competence proved to be determinants for engagement. Co-occurrence analysis reinforces that T1 and T2 form an inseparable core, indicating that content utility and delivery effectiveness are perceived as two sides of the same requirement. Protected time proved to be a transfer enabler, orbiting this core as a critical facilitator for the articulation between theory, practice, and context to materialize in the organizational environment. Institutional recognition sustains the continuity of development, considering that satisfied professionals tend to continue their education, while dissatisfied ones tend to avoid it \cite{birknerova2016differences}. Social interaction emerged as a substrate that enhances the other dimensions.

As a theoretical contribution, the study proposes a systemic view of organizational learning in technical contexts, articulating practical, pedagogical, structural, cultural, and social dimensions. As a practical contribution, it offers input for more contextualized training programs. The implementation of effective programs has the potential to improve organizational performance through higher productivity and lower turnover \cite{lee2023overcoming}.

Future research should prioritize longitudinal studies to assess the long-term impact of the identified dimensions \cite{lee2025overcoming}, cross-cultural comparisons, the role of emerging tools such as artificial intelligence, and the influence of sociodemographic variables on perceptions of corporate training.

\section*{Declaration on the Use of Artificial Intelligence}
 The authors used Generative AI tools (OpenAI ChatGPT and Google Gemini) exclusively for support in translating to the English language and for improving textual cohesion and clarity.

\bibliographystyle{sbc}
\bibliography{sbc-template}

\end{document}